\documentclass[prd,aps,nofootinbib,floatfix,11pt]{revtex4}
\usepackage{amsmath,graphicx,color,epsfig}

\begin{document}
\title{Hindered magnetic dipole transition in the covariant light-front approach}
\author{ Wei Wang
}
\affiliation{  Istituto Nazionale di Fisica Nucleare, Sezione di
Bari, Bari 70126, Italy }

\begin{abstract}
Hindered magnetic dipole transitions $\Upsilon(nS)\to \gamma
\eta_b(n'S)$ are studied in the covariant light-front approach.
Compared with the allowed magnetic dipole transitions, we find that
results for hindered magnetic dipole transitions are sensitive to
heavy quark mass and shape parameters of the light-front wave
functions. It is possible to tune the parameters so that the
predictions of branching fractions of
$\Upsilon(2S,3S)\to\gamma\eta_b$ are consistent with the recent
experimental data, but the relevant decay constant of $\eta_b$ is
much smaller than that of $\Upsilon(1S)$. We also generalize the
investigation to the charmonium sector  and find the the same
conclusion.
\end{abstract}

\maketitle

%


Since the photon energy in allowed magnetic dipole (M1) transitions
is limited, hindered M1 transitions with changes of principal
quantum numbers are believed to offer better chances to discover the
pseudoscalar quarkoniums such as $\eta_b$. Several years ago, the
CLEO collaboration set the upper bounds for this kind of
decays~\cite{Artuso:2004fp}
\begin{eqnarray}
 &&{\cal B }(\Upsilon(3S)\to\gamma\eta_b)<4.3\times 10^{-4},\nonumber\\
 &&{\cal B }(\Upsilon(2S)\to\gamma\eta_b)<5.1\times 10^{-4}.
\end{eqnarray}

In 2008 the BaBar collaboration observed a peak in the photon energy
spectrum at $E_\gamma=(921.2^{+2.1}_{-2.8}\pm2.4)$ MeV in radiative
$\Upsilon(3S)$ decay~\cite{:2008vj}. This is viewed as the first
observation of the $\eta_b$ meson which is the lowest bottomonium
ground state. The branching ratio of this radiative decay is
\begin{eqnarray}
 {\cal B }(\Upsilon(3S)\to\gamma\eta_b)&=&(4.8\pm0.5\pm1.2)\times 10^{-4}.
\end{eqnarray}
The measured mass
\begin{eqnarray}
 m(\eta_b)=(9388.9^{+3.1}_{-2.3}\pm2.7) {\rm MeV}\nonumber
\end{eqnarray}
gives the mass split \begin{eqnarray}
 \delta
 m_{\eta_b(1S)}=m(\Upsilon(1S))-m(\eta_b)=(71.4^{+2.3}_{-3.1}\pm2.7)
 {\rm MeV}.\end{eqnarray}
Subsequently the $\eta_b$ meson has also been observed in the
radiative $\Upsilon(2S)$ decay~\cite{:2009pz}
\begin{eqnarray}
 {\cal B }(\Upsilon(2S)\to\gamma\eta_b)&=&(3.9\pm1.1^{+1.1}_{-0.9})\times
 10^{-4}.
\end{eqnarray}
The ratio of the two branching fractions is measured as
\begin{eqnarray}
 R\equiv \frac{{\cal B }(\Upsilon(2S)\to\gamma\eta_b)}{{\cal B
 }(\Upsilon(3S)\to\gamma\eta_b)}&=&0.82\pm0.24^{+0.20}_{-0.19}.
\end{eqnarray}
The recently updated results by the CLEO
collaboration~\cite{Bonvicini:2009hs} are consistent with
measurements by the BaBar collaboration taking into account the
uncertainties
\begin{eqnarray}
 {\cal B }(\Upsilon(3S)\to\gamma\eta_b)&=&(7.1\pm1.8\pm1.1)\times
 10^{-4},\\
 {\cal B }(\Upsilon(2S)\to\gamma\eta_b)&<&8.4\times
 10^{-4}.
\end{eqnarray}

Decay widths of M1 transitions can be expressed in a well-known
formula
\begin{eqnarray}
 &&\Gamma(n^3S_1\to \gamma n'^1S_0)=\frac{4}{3}\alpha
 e_Q^2\frac{k^3}{m_Q^2}\left|\int_0^\infty dr r^2 R^*_{n'0}(r)j_0(\frac{kr}{2})R_{n0}(r)
 \right|^2.\label{eq:M1-space}
\end{eqnarray}
$\alpha=1/137$ is the fine-structure constant, $e_Q$ denotes the
charge in unit of $|e|$ of the transition quark, and  $m_Q$ denotes
the  quark mass. $k$ is the energy of the photon in the vector meson
rest frame.  For allowed transitions $(n=n')$, the emitted photon is
typically soft. One can expand the spherical Bessel function
$j_0(\frac{kr}{2})$ by $k$. Radiative corrections and
nonrelativistic corrections can be systematically studied in the
effective field theory~\cite{Brambilla:2005zw}. In particular the
predicted branching ratio of $J/\psi\to \gamma\eta_c$ is well
consistent with the experimental data.

The emitted photon in hindered M1 transitions is rather energetic
and the convergence of the expansion in $k$ becomes poor. On the
other aspect, the decay amplitude is zero at the leading power in
$k$ since the different wave functions are orthogonal. Higher power
terms contribute and they are sensitive to the treatments of the
nonperturbative dynamics in transition form factors. For more than
twenty years, hindered M1 transitions have received extensive
interests but theoretical predictions vary over several
orders~\cite{Godfrey:2001eb}. The recent experimental data could
offer an opportunity to investigate the dynamics in the quarkonia
transition and in particular might be helpful to constrain the form
of the wave functions.

In this work, we use the covariant light-front quark model (LFQM) to
investigate the M1 transitions and examine whether the commonly-used
light-front wave functions (LFWF) can simultaneously  explain the
experimental data for both allowed and hindered M1 transition. The
light-front QCD may be the only potential candidate to reconcile the
low energy quark model and the high energy parton model. The
LFQM~\cite{Jaus:1989au} can give a full treatment on spins of
hadrons using the so-called Melosh transformation. Physical
quantities are represented as the overlap of LFWF. These wave
functions are expressed in terms of the internal variables of the
quark and gluon degrees and thus are manifestly Lorentz invariant.
To preserve the covariance and remove the dependence of physical
quantities on the direction of the light-front, Jaus proposed the
covariant LFQM in which the zero-mode contributions are
systematically included~\cite{Jaus:1999zv}. The application of the
covariant LFQM to the s-wave and p-wave decay constants and form
factors is very
successful~\cite{Cheng:2003sm,Cheng:2004yj,Wang:2007sxa,Ke:2009ed,Ke:2009mn,Cheng:2009ms}.
Under this framework, the transition
$\Upsilon(1S)\to\gamma\eta_b(1S)$ has already been studied in
Ref.~\cite{Hwang:2006cua,Hwang:2008qi} (see also
Ref.~\cite{Choi:2007se}).

\begin{figure}
\includegraphics[scale=0.4]{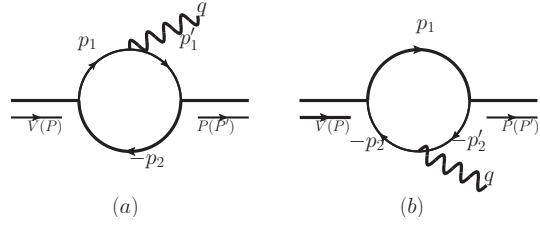}
\caption{Feynman diagrams for $V\to P\gamma$. }\label{fig:feyn}
\end{figure}

In the following we  employ the light-front decomposition of the
momentum $P^{}=(P^{
-}, P^{ +}, P^\prime_\bot)$, where $P^{\pm}=P^{0}\pm P^{3}$. 
These momenta can be expressed in
terms of the internal variables $(x_i, p_\bot)$ as:
 \begin{eqnarray}
 p_{1,2}^{+}=x_{1,2} P^{ +},
 p_{1,2\bot}=x_{1,2} P_\bot\pm p_\bot,
 \end{eqnarray}
with $x_1+x_2=1$. With these internal variables, one can define some
useful quantities
\begin{eqnarray}
 M^{2}_0
          &=&(e_1+e_2)^2=\frac{p^{2}_\bot+m_1^{2}}
                {x_1}+\frac{p^{2}_{\bot}+m_2^2}{x_2},\nonumber\\
 p_z&=&\frac{x_2 M_0}{2}-\frac{m_2^2+p^{2}_\bot}{2 x_2
 M_0},\;\;
 e^{}_i
          =\sqrt{m^{2}_i+p^{2}_\bot+p^{2}_z}.
 \end{eqnarray}
here $e_i$ can be interpreted as the energy of the quark/antiquark
and $M_0$ can be viewed as kinetic invariant mass of the meson
system. The transition amplitude of $V(P,\epsilon_V)\to
P\gamma^*(q,\epsilon_\gamma)$ is usually parametrized as
\begin{eqnarray}
 {\cal A}&=&
 ie\epsilon_{\mu\nu\rho\sigma}\epsilon^{*\mu}_\gamma\epsilon_V^\nu
 q^\rho P^\sigma V(q^2),
\end{eqnarray}
where the photon is firstly taken as off-shell $q^2\neq0$ and  the
convention $\epsilon^{0123}=1$ is adopted.
Feynman diagrams for this process are given in Fig.~\ref{fig:feyn},
in which the photon is emitted from the quark shown in (a) or from
the antiquark shown in (b). Due to the charge conjugation
invariance, these two diagrams provide identical contributions. It
is straightforward to evaluate these two diagrams:
\begin{eqnarray}
 {\cal A}&=& -iee_QN_c
 \int\frac{d^4p_1}{(2\pi)^4}\left\{\frac{H_VH_P'}{N_1N_2N_1'}s_{\mu\nu}^a
 +\frac{H_VH_P'}{N_1N_2N_2'}s_{\mu\nu}^b\right\}\epsilon_\gamma^{*\mu}\epsilon_V^\nu,
\end{eqnarray}
where $N_i= p_i^{2}-m_i^2$, $N_i'= p_i'^{2}-m_i^2$ and
\begin{eqnarray}
 s_{\mu\nu}^a&=&{\rm Tr}\left[\left(\gamma_\nu-\frac{p_{1\nu}-p_{2\nu}}{W_V}\right)(-p\!\!\!\slash_2+m_2)
 \gamma_5
 (p\!\!\!\slash'_1+m_1)\gamma_\mu(p\!\!\!\slash_1+m_1)\right],\\
 s_{\mu\nu}^b&=&{\rm
 Tr}\left[\left(\gamma_\nu-\frac{p_{1\nu}-p_{2\nu}}{W_V}\right)(-p\!\!\!\slash_2+m_2)
 \gamma_\mu(-p\!\!\!\slash'_2+m_2) \gamma_5(p\!\!\!\slash_1+m_1)\right].
\end{eqnarray}
Functions $H_V,H_P$, depending on the four momentum of the internal
quarks, are the wave functions for vector and pseudoscalar mesons.
In the absence of singularity in the $H_V,H_P$, integrating over the
minus component of the internal momentum will pick up the pole in
the propagators. Then the functions $H_{V,P}$ are replaced by
$h_{V,P}$ which only depends on the plus component and the
transverse momentum and $W_V$ is replaced by $w_V$
\begin{eqnarray}
 h_V&=& h_P=(M^2-M_0^2)\sqrt{\frac{x_1x_2}{N_c}}\frac{1}{\sqrt2
 M_0}\phi(x_2,p_\perp),\nonumber\\
 w_V&=&M_0+m_1+m_2,
\end{eqnarray}
where $\phi(x_2,p_\perp)$ denotes the momentum distribution inside
the meson. In order to preserve the covariance, it is suggested that
the zero mode contributions should be added in the covariant LFQM.
The inclusion of the zero-mode contribution corresponds to a proper
way to integrate out the minus component and after the integration
expressions for the form factor $V(q^2)$ are given as
follows~\cite{Hwang:2006cua,Hwang:2008qi}
\begin{eqnarray}
 V(q^2)&=&\frac{e_Q}{8\pi^3}\int dx_2
 d^2p_\perp\frac{2\phi_V(x_2,p_\perp)\phi_P(x_2,p_\perp')}{x_1M_0M_0'} \left(m_Q+\frac{2}{w_V}(p_\perp^2+\frac{(p_\perp\cdot
 q_\perp)^2}{q^2})\right).\label{eq:formfactor-LFQM}
\end{eqnarray}
and the decay width of $V\to P\gamma$ is then evaluated  as
\begin{eqnarray}
 \Gamma(V\to P\gamma)&=& \frac{4\pi\alpha}{3}\frac{(m_V^2-m_P^2)^3}{32\pi
 m_V^3}|V(0)|^2.
\end{eqnarray}

In the $m_Q\to\infty$ limit, the heavy quark inside the quarkonium
moves nonrelativistically. Integrating out the hard off-shell
degrees of freedom, one reaches an effective field theory known as
nonrelativistic QCD~\cite{Bodwin:1994jh} to deal with the low energy
dynamics. In a heavy quarkonium, the square of the transverse
momentum $p_\perp^2$ is of the order $\Lambda_{\rm QCD}^2$ and the
momentum fraction of the heavy quark $x_2\simeq 1/2$. The kinetic
invariant mass $(M_0,M_0^\prime)$ is roughly $2m_Q$. At the leading
power in $1/m_Q$, the form factor $V(0)$ in
Eq.~(\ref{eq:formfactor-LFQM}) is reduced to
\begin{eqnarray}
 V(0)&\simeq&\frac{e_Q}{8\pi^3}\int dx_2
 d^2p_\perp\frac{\phi_V(x_2,p_\perp)\phi_P(x_2,p_\perp')}{m_Q}.\label{eq:formfactor-LFQM-heavy-limit}
\end{eqnarray}
This formula, with the LFWF discussed in the following, could
formally reproduce the leading power behavior shown in
Eq.~(\ref{eq:M1-space}) in the momentum space.

For hindered M1 transition, the transition amplitude is from the
higher power corrections and one needs a specific approach to
calculate them. As we can see from the above discussion, the form
factors in the covariant LFQM are expressed as the overlap of the
LFWF. It contains the nonperturbaitve coupling of the quark pair
with a meson. Except for some limited cases, these quantities can
not be derived from the first principle. Practically one often
adopts a phenomenological form. The commonly-used wave function for
the $S$-wave mesons with lowest principle number is the
Gaussian-type
 \begin{eqnarray}
 \phi(x_2,p_{\bot})&=&4\left(\frac{\pi}{\beta^2}\right)^{3/4}\sqrt{\frac{dp_z}{dx_2}}~{\rm exp}\left(
  -\frac{p_{\bot}^2+p_z^2}{2\beta^2}
  \right),\label{eq:wavefn}\\
 \frac{dp_z}{dx_2}&=&\frac{e_1e_2}{x_1x_2 M_0}.\nonumber
 \end{eqnarray}
In the case of the wave functions of the $2S,3S$ mesons, we adopt
the solutions of the harmonic-oscillator problem
\begin{eqnarray}
 \phi_{2S}(r)&=& \sqrt
 {\frac{1}{6}}\left(\frac{\beta^2}{\pi}\right)^{3/4}\left(\frac{3}{4}-\frac{\beta^2 r^2}{2}\right)
  {\rm exp}\left(-\frac{\beta^2 r^2}{2}\right),\nonumber\\
 \phi_{3S}(r)&=& \sqrt
 {\frac{2}{15}}\left(\frac{\beta^2}{\pi}\right)^{3/4}{\rm exp}\left(-\frac{\beta^2
 r^2}{2}\right) \times
 \left(\frac{15}{4}-5\beta^2r^2+\beta^4 r^4\right).
 \end{eqnarray}
The Fourier transformations, with the Jacobi determinant
$\sqrt{\frac{dp_z}{dx_2}}$, are derived as
\begin{eqnarray}
 \phi_{2S}(x_2,p_\perp)&=& 4\sqrt
 {\frac{2}{3}}\left(\frac{\pi}{\beta^2}\right)^{3/4}
 \sqrt{\frac{dp_z}{dx_2}}{\rm exp}\left(-\frac{p_{\bot}^2+p_z^2}{2\beta^2}\right) \times\left(\frac{p_{\bot}^2+p_z^2}{\beta^2}-\frac{3}{2}\right),\nonumber\\
 \phi_{3S}(x_2,p_\perp)&=&  4\sqrt
 {\frac{2}{15}}\left(\frac{\pi}{\beta^2}\right)^{3/4}
  \sqrt{\frac{dp_z}{dx_2}}{\rm
  exp}\left(-\frac{p_{\bot}^2+p_z^2}{2\beta^2}\right)
  \left(\frac{(p_{\bot}^2+p_z^2)^2}{\beta^4}-\frac{5(p_{\bot}^2+p_z^2)}{\beta^2}+\frac{15}{4}\right).
\end{eqnarray}
This kind of LFWF for the $2S$ mesons has been used in
Ref.~\cite{Hwang:2008qi} while its light-cone form is studied in
Ref.~\cite{Braguta:2007tq}. It would be meaningful to use the
allowed and hindered M1 transitions to constrain the forms of the
LFWF and in particular to test the harmonic-oscillator wave
functions given above, as most of the previous studies in the
covariant LFQM mainly focus on mesons without any radial excitation.

%

\begin{table}
\caption{Input parameters in the light-front wave functions (in
units of GeV). }\label{tab:inputs}
\begin{tabular}{ccccc}
 \hline
  &  $\Upsilon(1S)$       & $\Upsilon(2S)$             & $\Upsilon(3S)$      \\
 \hline
 Mass &  9.46030   & 10.02326     & 10.3552\\
 \hline
 $f_\Upsilon$  & $0.7150\pm0.0048$     & $0.4974\pm0.0045$      & $0.4301\pm0.0039$ \\
 \hline
 $\beta$ &
  $1.323^{+0.037}_{-0.035} $ & $0.940^{+0.026}_{-0.024} $
 &  $0.808^{+0.022}_{-0.020}$\\
 \hline
  &  $J/\psi$       & $\psi(2S)$                    \\
 \hline
 Mass &  3.069616  & 3.68609      \\
 \hline
 $f_\psi$  & $0.4163\pm0.0053$     & $0.2961\pm0.0025$        \\
 \hline
 $\beta$   &    $0.641^{+0.027}_{-0.025}$
  &  $0.497^{+0.019}_{-0.017}$      \\
 \hline
\end{tabular}
\end{table}

Two kinds of inputs are required in the numerical analysis: masses
(for hadrons and constituent quarks) and the shape parameters in the
wave functions. For the heavy quarks, the difference between the
current mass and the constituent mass is small. From the
PDG~\cite{Amsler:2008zzb}, the current masses are
$m_c=(1.27^{+0.07}_{-0.11})$ GeV and $m_b=(4.20^{+0.17}_{-0.07})$
GeV in the ${\rm \overline{MS}}$ renormalization scheme. In this
work, we will choose three different values for the constituent
quark masses
 \begin{eqnarray}
  m_c=(1.1,1.3,1.5) ~{\rm GeV},\;\;    m_b=(4.0,4.4,4.8) {\rm GeV}.
 \end{eqnarray}
As for hadron masses, all of them used in the present analysis have
been measured except the one of $\eta_b(2S)$. We will use the mass
split in the charm sector~\cite{Amsler:2008zzb}
\begin{eqnarray}
 \delta m_{\eta_c(1S)}=116.5 {\rm MeV},\;\;\;\delta m_{\eta_c(2S)}=48.1 {\rm MeV}\nonumber
\end{eqnarray}
to estimate the mass split for the $\eta_b(2S)$
\begin{eqnarray}
 \delta m_{\eta_b(2S)}=\frac{\delta m_{\eta_b(1S)}}{\delta m_{\eta_c(1S)}}\delta m_{\eta_c(2S)}=29.5{\rm MeV}.
\end{eqnarray}
Shape parameters of the LFWF are usually determined by decay
constants of hadrons whose expressions are given in
Ref.~\cite{Cheng:2003sm}. Decay constants of $\Upsilon(nS)$ are
extracted from the partial decay width of the leptonic $\Upsilon\to
e^+e^-$ decays
\begin{eqnarray}
\Gamma_{ee}\equiv\Gamma(\Upsilon\to e^+e^-)=\frac{4\pi \alpha^2e_b^2
f_{\Upsilon}^2}{3m_{\Upsilon}}.
\end{eqnarray}
The  shape parameters (corresponding to $m_b=4.4\pm0.4$ GeV and
$m_c=1.3\pm0.2$ GeV) are collected in table~\ref{tab:inputs}. In the
case of $\eta_b$ mesons, their decay constants are not measured at
present. In the future, these decays may not be well constrained as
the uncertainties in the ideal mode $\eta_b\to2\gamma$ would be
typically large. In this work, we first try to tune the shape
parameter so that the predictions of branching fractions of
$\Upsilon(2S,3S)\to\gamma\eta_b$ decays are consistent with the data
provided by the BaBar collaboration.


With above parameters, results for the branching ratios in
$\Upsilon$ decays are collected in table~\ref{tab:results-etab} and
table~~\ref{tab:results-etab-2}, in which we have used three sets
inputs for the shape parameters and three different values for the
quark mass. Several remarks are give in order. Firstly, from these
tables one can see that the decay constant of the $\eta_b$ is not
very sensitive to the shape parameter and the quark mass $m_b$.
Secondly the allowed channels $\Upsilon(nS)\to \eta_b(nS)(n=1,2)$
are not sensitive to these inputs either. On the contrary, results
for hindered channels are strongly dependent on the input
parameters, which is not beyond expectation.  If we adopt the
$m_b=4.4$ GeV and $\beta_{\eta_b}=0.93$ GeV, theoretical predictions
of branching fractions of $\Upsilon(2S,3S)\to\gamma\eta_b$ are
consistent with the experimental data given in
Ref.~\cite{:2008vj,:2009pz}. The ratio of branching fractions is
predicted as $0.66$.

In the heavy quark limit, pseudoscalar mesons and vector mesons
belong to the same supermultiplet and their decay constants are
almost the same. If we adopt the same decay constant for $\eta_b$
with its vector partner, the shape parameter is determined as
$\beta_{\eta_b}=1.451$ GeV and theoretical predictions of branching
ratios are given as
\begin{eqnarray}
 {\cal B}(\Upsilon(1S)\to\gamma\eta_b)&=&3.2\times 10^{-4},\nonumber\\
 {\cal B}(\Upsilon(2S)\to\gamma\eta_b)&=&5.8\%,\nonumber\\
 {\cal B}(\Upsilon(3S)\to\gamma\eta_b)&=&13.7\%.
\end{eqnarray}
The latter two results are much larger than the experimental data
given earlier. These inconsistencies between theoretical results and
the data may imply either the inappropriate form of the LFWF used in
this analysis, or a small decay constant for $\eta_b$ compared with
that for $\Upsilon$.

\begin{table}
\caption{Branching ratios (in units of $10^{-4}$) of M1 transitions:
$\Upsilon(nS)\to\gamma\eta_b(n'S)$. The $b$ quark mass is used as
$m_b=4.4$ GeV. }\label{tab:results-etab}
\begin{tabular}{cccc ccccc}
 \hline
  $\beta_{\eta_b}$  &  0.90& 0.93 &0.96   \\
 \hline
 $f_{\eta_b}$& 389& 406 & 424 \\
 \hline
 ${\cal B}(\Upsilon(1S)\to\eta_b)$         &  2.7
   &     $2.8$   &   2.9
 \\
 \hline
 ${\cal B}(\Upsilon(2S)\to\eta_b)$
 & 21.3
 &   5.0  &  $0.003$
 \\
 \hline
 ${\cal B}(\Upsilon(3S)\to\eta_b)$      &   2.3  &   7.7  &  18.6
 \\
 \hline
 $R$ & 9.4  & 0.66& $1.5\times 10^{-4}$
 \\
 \hline\hline $\beta_{\eta_b(2S)}$  &  0.75& 0.8 &0.85  \\
 \hline
 $f_{\eta_b(2S)}$& 343& 372 & 400    \\
 \hline
 ${\cal B}(\Upsilon(2S)\to\eta_b(2S))$       & 0.3  &  0.3 &  0.4
 \\
 \hline
 ${\cal B}(\Upsilon(3S)\to\eta_b(2S))$       &  40.3  &   3.1  &
 6.7
 \\
 \hline\hline
\end{tabular}
\end{table}

\begin{table}
\caption{Branching ratios (in units of $10^{-4}$) of M1 transitions:
$\Upsilon(nS)\to\gamma\eta_b(n'S)$. The shape parameters of $\eta_b$
and $\eta_b(2S)$ are used as $\beta_{\eta_b}=0.93$ GeV;
$\beta_{\eta_b(2S)}=0.80$ GeV. }\label{tab:results-etab-2}
\begin{tabular}{cccc ccccc}
 \hline
  $m_b$    &4.0 &4.4  &4.8\\
 \hline
 $f_{\eta_b}$ &   $419$& 406 &  394\\
 \hline
 ${\cal B}(\Upsilon(1S)\to\eta_b)$
 &   $3.3$
   &     $2.8$
 &   2.5
 \\
 \hline
 ${\cal B}(\Upsilon(2S)\to\eta_b)$
 &    $24.8$
 &   5.0
 &   0.003
 \\
 \hline
 ${\cal B}(\Upsilon(3S)\to\eta_b)$
 &   $3.3$   &   7.7
 &    14.4
 \\
 \hline
 $R$     & 7.5 & 0.66 & $0.0002$
 \\
 \hline\hline
   $m_b$    &4.0 &4.4 &4.8\\
 \hline
 $f_{\eta_b(2S)}$    & 379 & 372   &364\\
 \hline
 ${\cal B}(\Upsilon(2S)\to\eta_b(2S))$          &  0.4 &  0.3   & 0.3
 \\
 \hline
 ${\cal B}(\Upsilon(3S)\to\eta_b(2S))$         &   18.2   &   3.1&  0.055
 \\
 \hline
\end{tabular}
\end{table}

\begin{table}
\caption{Results  for the branching ratios of M1 transitions
involving $\eta_c$ in the covariant LFQM (in units of $10^{-2}$ for
$J/\psi \to \eta_c$ and $10^{-3}$ for $\psi(2S) \to \eta_c$). In the
left sector, the mass is used as $m_c=1.3$ GeV; while in the right
sector, the shape parameter is $\beta_{\eta_c}=0.55$ GeV.
}\label{tab:results-etac}
\begin{tabular}{cccc||ccccc}
 \hline
 $\beta_{\eta_c}$ &  0.5& 0.55 &0.6 & $m_c$  &1.1  &1.3 &1.5\\
 \hline
 $f_{\eta_c}$& 253 & 281& 308 & $f_{\eta_c}$&    283&  281 &276\\\hline
  ${\cal B}(J/\psi \to \eta_c)$   &   3.0 &   3.1  &   3.1
 &   ${\cal B}(J/\psi \to \eta_c)$ &  $3.8$ &3.1
 &  2.5
 \\
 \hline
 ${\cal B}(\psi(2S) \to \eta_c)$  &  5.5 &   4.2  &  33.4 & ${\cal B}(\psi(2S) \to \eta_c)$  &  $0.6$ &4.2 &
 15.3
 \\
 \hline
 \hline%
\end{tabular}
\end{table}

As another check, we will extend the above discussion into the
charmonium sector: $\psi(nS)\to \gamma \eta_c(n'S)$.  The relevant
experimental data is given as~\cite{Amsler:2008zzb,Gao:2009ad}
\begin{eqnarray}
 {\cal B}(J/\psi\to\gamma\eta_c) &=& (1.7\pm0.4)\%,\\
 {\cal B}(\psi(2S)\to\gamma\eta_c) &=& (3.4\pm0.5)\times 10^{-3},\\
 {\cal B}(\psi(2S)\to\gamma\eta_c(2S)) &<& 7.4\times 10^{-4}.
\end{eqnarray}
Again with proper inputs the results in table~\ref{tab:results-etac}
could be consistent with these data, but the relevant decay constant
of $\eta_c$ is much smaller that of its vector partner $J/\psi$. Our
result for ${\cal B}(\psi(2S)\to \eta_c(2S))$
\begin{eqnarray}
 {\cal B}(\psi(2S)\to\gamma\eta_c(2S)) &=&(6.3^{+0.2+1.2}_{-1.2-1.0})\times 10^{-4}.
\end{eqnarray}
is approaching the experimental upper bound. The first uncertainties
are from $m_c=(1.3\pm0.2)$GeV and the second uncertainties are from
$\beta_{\eta_c(2S)}=(0.45\pm0.05)$ GeV.
In Ref.~\cite{Hwang:2008qi}, the inputs $m_c=1.56$ GeV,
$\beta_{\eta_c}=0.820$ GeV and $\beta_{J/\psi}=0.613$ GeV are
adopted and prediction of ${\cal B}(J/\psi(1S)\to\gamma\eta_c)$ is
consistent with the data reported by the CLEO
collaboration~\cite{:2008fb}. If we adopt these parameters, the
branching ratio of $\psi(2S)\to\gamma\eta_c$ is larger than the
experimental data by two orders
\begin{eqnarray}
 {\cal B}( \psi(2S)\to\gamma\eta_c)&=& 18.2\%.
\end{eqnarray}
Inspired by these inconsistencies, one may get the same conclusion
as in the bottomonium sector.


In conclusion,  we studied the hindered M1 transitions
$\psi(2S)\to\gamma\eta_c$ and the
$\Upsilon(nS)\to\gamma\eta_b(1S,2S)$ in the covariant light-front
quark model. Compared with the allowed M1 transitions, theoretical
results for hindered channels are found to strongly depend on  quark
masses and the shape of the light-front wave functions. Because of
this sensitivity, these transitions offer  good laboratories to
study the dynamics between the transitions and constrain the forms
of the light-front wave functions.  Using the harmonic-oscillator
wave functions and the experimental data, our results show that the
decay constants of the pseudoscalar mesons are required to be much
smaller than the decay constants of their vector partners. With the
precise decay constants measured in various other channels in the
future, different forms of the wave functions could be directly
tested in the channels studied here. A more practical way in the
future is to incorporate more radiative transitions between heavy
quarkonium states as in Ref.~\cite{DeFazio:2008xq}.

\emph{Note added}: During the long-time revision of this work, a
similar paper~\cite{Ke:2010tk} appears. One of the important
differences is: we have shown that hindered M1 transitions are
sensitive to the shape of the wave functions as they should.

\section*{Acknowledgements}

I would like to acknowledge Y. Jia and C.D. L\"u for fruitful
discussions and kind encouragement, H.Y. Cheng, C.K. Chua and X.H.
Liu for useful discussions on the harmonic oscillator wave
functions, Y.M. Wang for reading the manuscript and sharing the
results of $\Upsilon(3S)\to\gamma\eta_b$ in light-cone sum rules,
and Z.T. Wei for the collaboration in the light-front quark model
and the comment on the sensitivity to light-front wave functions.


\begin{thebibliography}{99}


\bibitem{Artuso:2004fp}
  M.~Artuso {\it et al.}  [CLEO Collaboration],
  Phys.\ Rev.\ Lett.\  {\bf 94}, 032001 (2005).


\bibitem{:2008vj}
  B.~Aubert {\it et al.}  [BABAR Collaboration],
  Phys.\ Rev.\ Lett.\  {\bf 101}, 071801 (2008)
  [Erratum-ibid.\  {\bf 102}, 029901 (2009)].




\bibitem{:2009pz}
  B.~Aubert {\it et al.}  [BABAR Collaboration],
  Phys.\ Rev.\ Lett.\  {\bf 103}, 161801 (2009)
  [arXiv:0903.1124 [hep-ex]].




\bibitem{Bonvicini:2009hs}
   G.~Bonvicini {\it et al.}  [The CLEO Collaboration],
  arXiv:0909.5474 [hep-ex].


\bibitem{Brambilla:2005zw}
  N.~Brambilla, Y.~Jia and A.~Vairo,
  Phys.\ Rev.\  D {\bf 73}, 054005 (2006).



\bibitem{Godfrey:2001eb}
  S.~Godfrey and J.~L.~Rosner,
  Phys.\ Rev.\  D {\bf 64}, 074011 (2001)
  [Erratum-ibid.\  D {\bf 65}, 039901 (2002)]; and many references
  therein;
For another new insight into hindered magnetic dipole transition,
see:
  Y.~Jia, J.~Xu and J.~Zhang,
  arXiv:0901.4021 [hep-ph].

%



\bibitem{Jaus:1989au}
  W.~Jaus,
  Phys.\ Rev.\  D {\bf 41}, 3394 (1990);
  Phys.\ Rev.\  D {\bf 44}, 2851 (1991);
  H.~Y.~Cheng, C.~Y.~Cheung and C.~W.~Hwang,
  Phys.\ Rev.\  D {\bf 55}, 1559 (1997);
  H.~M.~Choi, C.~R.~Ji and L.~S.~Kisslinger,
  Phys.\ Rev.\  D {\bf 65}, 074032 (2002).




\bibitem{Jaus:1999zv}
  W.~Jaus,
  Phys.\ Rev.\  D {\bf 60}, 054026 (1999).


\bibitem{Cheng:2003sm}
  H.~Y.~Cheng, C.~K.~Chua and C.~W.~Hwang,
  Phys.\ Rev.\  D {\bf 69}, 074025 (2004).



\bibitem{Cheng:2004yj}
  H.~Y.~Cheng and C.~K.~Chua,
  Phys.\ Rev.\  D {\bf 69}, 094007 (2004)
  [arXiv:hep-ph/0401141].

\bibitem{Wang:2007sxa}
  C.~D.~Lu, W.~Wang and Z.~T.~Wei,
  Phys.\ Rev.\  D {\bf 76}, 014013 (2007)
  [arXiv:hep-ph/0701265];
  W.~Wang, Y.~L.~Shen and C.~D.~Lu,
  Eur.\ Phys.\ J.\  C {\bf 51}, 841 (2007)
  [arXiv:0704.2493 [hep-ph]];
  W.~Wang and Y.~L.~Shen,
  Phys.\ Rev.\  D {\bf 78}, 054002 (2008);
  W.~Wang, Y.~L.~Shen and C.~D.~Lu,
  Phys.\ Rev.\  D {\bf 79}, 054012 (2009)
  [arXiv:0811.3748 [hep-ph]];
  X.~X.~Wang, W.~Wang and C.~D.~Lu,
  Phys.\ Rev.\  D {\bf 79}, 114018 (2009)
  [arXiv:0901.1934 [hep-ph]];
  C.~H.~Chen, Y.~L.~Shen and W.~Wang,
  arXiv:0911.2875 [hep-ph].

\bibitem{Ke:2009ed}
  H.~W.~Ke, X.~Q.~Li and Z.~T.~Wei,
  Phys.\ Rev.\  D {\bf 80}, 074030 (2009)
  [arXiv:0907.5465 [hep-ph]].

\bibitem{Ke:2009mn}
  H.~W.~Ke, X.~Q.~Li and Z.~T.~Wei,
  arXiv:0912.4094 [hep-ph].

\bibitem{Cheng:2009ms}
  H.~Y.~Cheng and C.~K.~Chua,
  arXiv:0909.4627 [hep-ph].


\bibitem{Hwang:2006cua}
  C.~W.~Hwang and Z.~T.~Wei,
  J.\ Phys.\ G {\bf 34}, 687 (2007).

\bibitem{Hwang:2008qi}
  C.~W.~Hwang,
  Eur.\ Phys.\ J.\  C {\bf 62}, 499 (2009).



\bibitem{Choi:2007se}
  H.~M.~Choi,
  Phys.\ Rev.\  D {\bf 75}, 073016 (2007).






\bibitem{Bodwin:1994jh}
  G.~T.~Bodwin, E.~Braaten and G.~P.~Lepage,
  Phys.\ Rev.\  D {\bf 51}, 1125 (1995)
  [Erratum-ibid.\  D {\bf 55}, 5853 (1997)].





\bibitem{Braguta:2007tq}
  V.~V.~Braguta,
  Phys.\ Rev.\  D {\bf 77}, 034026 (2008).

\bibitem{Amsler:2008zzb}
  C.~Amsler {\it et al.}  [Particle Data Group],
  Phys.\ Lett.\  B {\bf 667}, 1 (2008).


%



\bibitem{Gao:2009ad}
  D.~Cronin-Hennessy  [CLEO Collaboration],
  arXiv:0910.1324 [hep-ex].



\bibitem{:2008fb}
  R.~E.~Mitchell {\it et al.}  [CLEO Collaboration],
  Phys.\ Rev.\ Lett.\  {\bf 102}, 011801 (2009).







\bibitem{DeFazio:2008xq}
  F.~De Fazio,
  Phys.\ Rev.\  D {\bf 79}, 054015 (2009).



\bibitem{Ke:2010tk}
  H.~W.~Ke, X.~Q.~Li and X.~Liu,
  arXiv:1002.1187 [hep-ph].

\end{thebibliography}
\end{document}